\documentstyle[prl,aps]{revtex}
\begin{document}
\draft
\twocolumn[\hsize\textwidth\columnwidth\hsize\csname @twocolumnfalse\endcsname

\title{Singularity of the density of states in the
two-dimensional Hubbard model
from finite size
scaling of Yang-Lee zeros}

\author{E. Abraham$^1$,
I.M. Barbour$^2$,
P.H. Cullen$^1$,
E.G. Klepfish$^3$ , E.R. Pike$^3$ and Sarben Sarkar$^3$}
\address{$^1$Department of Physics, Heriot-Watt University, Edinburgh
EH14 4AS, UK\\
$^2$Department of
Physics, University of Glasgow, Glasgow G12 8QQ, UK\\
$^3$Department of Physics, King's College London, London WC2R 2LS, UK}

\date{\today}
\maketitle

\begin{abstract}
A finite size scaling is applied to the Yang-Lee zeros
of the grand canonical partition function for the 2-D Hubbard model
in the complex chemical potential plane. The logarithmic scaling
of the imaginary part of the zeros with the system size indicates
a singular dependence of the carrier density on the chemical potential.
Our analysis points to a second-order phase transition
with critical exponent ${1\over \delta}={1\over 2}\pm {1\over 3}$
which leads to a divergence of the electronic susceptibility.
We
interprete these results as reflecting singular behaviour of
the density of states in the quasiparticle spectrum.
\end{abstract}

\vskip2pc]
\narrowtext

\section{Introduction}
Quantum Monte Carlo techniques are widely applied in the investigation
of the phase structure of the Hubbard model in an attempt to
explore its relationship to high temperature superconductivity
(HTSC).
The pioneering works of Blankenbecler, Scalapino and
Sugar [\ref{Blankenbecler}]
and Hirsch [\ref{Hirsch}] outlined the finite-temperature
Quantum Monte Carlo algorithms for treating
systems of interacting fermions, thus allowing efficient numerical
simulations of the model. These simulations showed remarkable
stability for the half-filled model with on-site Coulomb repulsion,
and could be extended to the doped case in the relatively high temperature
regime.
However, in the low-temperature and finite-doping regime, which is more
relevant to HTSC, the simulations become less
reliable due to the frequent
appearance of field configurations with negative
statistical weight, the so-called sign problem.
This problem severely obscures numerical evaluation of thermodynamical
averages of correlation functions and local condensates with
subsequent difficulties in establishing the phase structure of
the Hubbard model.

Barbour and Klepfish [\ref{BarbourKlepfish}] explored an alternative
method of detecting a phase transition in the Hubbard model.
This method is based on the application of the general
ideas of the Yang-Lee theorem [\ref{YangLee}] regarding
the distribution of zeros of the grand canonical partition function in the
complex fugacity plane.
The theorem was originally proved
for the Ising model and lattice
gas[\ref{YangLee}]. Later extensions of this theorem
include the case of higher-order Ising spins[\ref{Griffiths}], Ising
models with multiple spin interactions, the quantum Heisenberg
model[\ref{SuzukiFisher}], the classical XY and Heisenberg models
[\ref{DunlopNewman}], and some continuous spin systems[\ref{SimonGriffiths}].
Recently, a generalization of the Yang-Lee theorem to the
asymmetric first-order transition was made[\ref{KCLee}].

Following the Yang and Lee treatment, one identifies
a phase transition by the appearance of zeros of the
partition function in the complex
chemical potential ($\mu$) plane near the real
axis, namely the physical region. For any finite system simulated
on a lattice, these zeros will never be real. However,
one may identify a group of zeros
in the vicinity of the real axis which, if there is a phase
transition, converge towards it
as the system volume grows. The critical value of the chemical
potential is thus the thermodynamic limit of the
zero with the smallest imaginary part.

The $4^2$ lattice simulations presented in Ref.[\ref{BarbourKlepfish}]
suggested that, from the scaling of the smallest zeros
in the complex $\mu$ plane with the inverse temperature $\beta$,
the two-dimensional single-band Hubbard model
undergoes a phase transition only at zero temperature.
This conclusion is in principle
supported by the Mermin-Wagner theorem which states
the absence of the off-diagonal long range order in two-dimensional
systems at temperatures above zero [\ref{MerminWagner}].
However, it is not clear whether this theorem excludes any
kind of phase transition in a doped two-dimensional Hubbard
model. Therefore a thorough finite size analysis is required
to validate or reject the above conclusion, as was pointed
out in [\ref{BarbourKlepfish}].

The distribution of zeros in the vicinity of the real axis,
namely the angle of the zeros' locus and the
scaling of the location of these zeros in the complex plane,
allows one to derive the critical exponents for the studied
system [\ref{ItzyksonPearsonZuber}],
[\ref{Privman}], [\ref{Kenna}].
This derivation is based upon the scaling hypothesis
which is fundamental to the renormalization-group
approach. In this work we apply the ideas developed in statistical
mechanics models to investigate the criticalities in the Hubbard
model of strongly correlated electrons.

One such criticality manifests itself through
the macroscopic density of carriers.
When the latter is derived by differentiating
the free energy with respect to the chemical potential, it can serve
as an indicator of a phase transition controlled by
the chemical potential. As in order-disorder transitions, one would expect
a symmetry breaking signalled by an order parameter. In this model, the
particle-hole symmetry is broken by introducing an ``external field''
which causes the particle density to become non-zero.
Furthermore, the possibility of the free energy
having a singularity at some finite value of
the chemical potential is not excluded: in fact it can
be a transition indicated by a divergence of
the correlation length. A singularity of the free energy at
finite ``external
field" was found in finite-temperature
lattice QCD by using the Yang-Lee analysis for the chiral
phase transition [\ref{BarbourBellKlepfish}].
A possible scenario for such a transition at finite
chemical potential, is one in which
the particle density consists of two
components derived from the regular and singular parts
of the free energy.

Since we are dealing with a grand
canonical ensemble, the particle number can be calculated
for a given chemical potential as opposed to constraining the
chemical potential
by a fixed particle number. Hence the
chemical potential can be thought of as an external field
for exploring the behaviour of the free energy.
{}From the microscopic point of view, the critical values of the
chemical potential are associated with singularities of
the density of states. Transitions
related to the singularity
of the density of states are known as Lifshitz transitions
[\ref{Lifshitz}]. In metals these transitions only
take place at zero temperature, while at
finite temperatures the singularities are rounded. However, for a small
ratio of temperature to the deviation
from the critical values of the chemical potential, the singularity
can be traced even at finite temperature. Lifshitz transitions
may result from topological changes of the Fermi surface, and
may occur inside the Brillouin zone as well as on its boundaries
[\ref{Abrikosov}].
In the case of strongly correlated electron systems
the shape of the Fermi surface is indeed affected,
which in turn may lead to an extension of the Lifshitz-type
singularities into the finite-temperature regime.

In relating the
macroscopic quantity of the carrier density
to the density of quasiparticle states, we assumed
the validity of a single particle excitation picture.
Whether strong correlations completely distort this description
is beyond the scope of the current study. However, the identification
of the criticality using the Yang-Lee analysis, remains
valid even if collective excitations prevail.

The paper is organised as follows.
In Section 2 we outline the essentials of
the computational technique used to simulate the grand canonical
partition function and present its expansion as a polynomial
in the fugacity variable. In Section 3 we present the Yang-Lee zeros
of the partition function calculated on
$6^2$--$10^2$ lattices and highlight their
qualitative differences from the $4^2$ lattice.
In Section 4 we analyse the finite size scaling of the Yang-Lee zeros
and compare it to the real-space renormalization group prediction
for a second-order phase transition.
Finally, in Section 5 we present a
summary of our results and an outlook for
future work.

\section{Simulation algorithm and fugacity expansion of the grand canonical
partition function}
The model we are studying in this work is a two-dimensional
single-band Hubbard Hamiltonian
\begin{eqnarray}\label{Hubbard}
{\hat H}&=&-t\sum_{<i,j>,\sigma}c_{i,\sigma}^\dagger c_{j,\sigma}+
U\sum_{i}\left(n_{i+}-{1\over 2}\right)\left(n_{i-}-{1\over 2}\right)
\nonumber\\
&-&\mu\sum_{i}(n_{i+}+
n_{i-})
\end{eqnarray}
where the $i,j$ denote the nearest neighbour spatial
lattice sites, $\sigma$ is
the spin degree of
freedom and $n_{i\sigma}$ is the electron number operator
$c_{i\sigma}^\dagger c_{i\sigma}$. The constants $t$ and $U$ correspond
to the hopping parameter and the on-site Coulomb repulsion respectively.
The chemical potential $\mu$ is introduced such that $\mu=0$
corresponds to half-filling, {\it i.e.} the actual chemical potential
is shifted from $\mu$ to $\mu - {U\over 2}$.

The finite-temperature grand canonical
partition function is given by
\begin{equation}\label{partition}
Z={\rm Tr}\left(e^{-\beta \hat H}\right).
\end{equation}
Following
Hirsch [\ref{Hirsch}] and White {\it et al.} [\ref{White}] we rewrite
$Z$ as
\begin{equation}\label{partone}
Z={\rm Tr}\prod_{l=1}^{n_\tau}\left(e^{-dt\hat V}
e^{-dt\hat K}e^{dt \mu \hat N}\right)
\end{equation}
where $\hat K$ corresponds to the nearest neighbour
hopping term in the Hubbard
hamiltonian (\ref{Hubbard}) and
$\hat V$ to the on-site interaction
including quartic products of fermion fields, and $\hat N$ to the
particle number operator.
This decomposition, based on the Trotter
formula [\ref{Suzuki}],
introduces a systematic error proportional
to $dt^2$.
Thus $Z$ can be represented as a path integral of a 2+1-dimensional system
where the range of the additional dimension (imaginary time) corresponds
to the inverse temperature via
$\beta=dt\times n_\tau$.
The quartic interaction can be re-written in terms of
Ising fields $s_{i,l}$ using the discrete HS
transformation [\ref{Hirsch}]
\begin{equation}\label{Vnew}
e^{-dt \hat V}=e^{-{{dtU}\over 4}}{1\over 2}\sum_{s_{i,l}=\pm 1}
e^{-dts_{i,l}\gamma(n_{i+}-n_{i-})}
\end{equation}
where $i,l$ is the space-time index of a lattice site and
the coupling $\gamma$ is related to the original on-site
repulsion constant by
\begin{equation}
\cosh{(dt \gamma)}=\exp{\left({{dt U}\over2}\right)}.
\end{equation}

This transformation enables one to integrate out
the fermionic degrees of freedom and the resulting partition
function is written as an ensemble average of
a product of two determinants
\begin{equation}\label{parttwo}
Z=
\sum_{\{s_{i,l}=\pm 1\}} \tilde z =
\sum_{\{s_{i,l}=\pm 1\}} \det(M^+)\det(M^-)
\end{equation}
such that
\begin{equation}\label{defM}
M^\pm = \left(I+P^\pm\right)=\left(I+\prod_{l=1}^{n_\tau}B_l^{\pm}\right)
\end{equation}
where
the matrices $B_l^\pm $ are
defined as
\begin{equation}\label{Bls}
B_l^\pm=e^{-(\pm dt V)}e^{-dtK}e^{dt \mu}
\end{equation}
with $V_{ij}=
\delta_{ij}s_{i,l}$ and $K_{ij}=1$ if
$i,j$ are nearest neighbours and $K_{ij}=0$ otherwise.
The matrices in (\ref{defM}) and (\ref{Bls}) are of size
$(n_xn_y)\times(n_xn_y)$,
corresponding to the spatial size of the lattice.

The expectation value of a physical
observable at chemical potential
$\mu$,
$<O>_\mu$, is given
by
\begin{equation}\label{observa}
<O>_\mu = {{\int O \tilde z(\mu)}\over {\int \tilde z(\mu)}}\equiv
{{\sum\limits_{\{s_{i,l=\pm 1}\}}O(\{s_{i,l}\})\tilde z(\mu,\{s_{i,l}\})}\over
{\sum\limits_{\{s_{i,l=\pm 1}\}}\tilde z(\mu,\{s_{i,l}\})}}
\end{equation}
where the sum over the configurations of Ising fields
is denoted by an integral.
Since $\tilde z(\mu)$ is not positive definite for Re$(\mu) \neq 0$
we weight the ensemble of configurations by the absolute
value of $\tilde z(\mu)$ at some $\mu = \mu_0$. Thus
\begin{equation}\label{observb}
<O>_\mu =
{{\int{{O \tilde z(\mu)}\over{|\tilde z(\mu_0)|}}
|\tilde z(\mu_0)|}\over
{\int {{\tilde z(\mu)}\over {|\tilde z(\mu_0)|}}
|\tilde z(\mu_0)|}}
={{\left<{{O\tilde z(\mu)}\over {|\tilde z(\mu_0)|}}\right>_{\mu_0}}\over
{\left<{{\tilde z(\mu)}
\over {|\tilde z(\mu_0)|}}\right>_{\mu_0}}}
\end{equation}
The partition function $Z(\mu)$ is given by
\begin{equation}\label{newzz}
Z(\mu)\propto\left<{{\tilde z(\mu)}\over {|\tilde z(\mu_0)|}}\right>_{\mu_0}.
\end{equation}
The angular brackets notation ($<>_{\mu_0}$) is to be understood
here and in what follows as Monte Carlo integration with importance sampling
based on the absolute value of the $\tilde z$
calculated at an update chemical potential
$\mu_0$.
The normalization of $Z$
is irrelevant as can be seen from eqn.(\ref{observb}).

It was shown in [\ref{BarbourKlepfish}] that a unitary transformation
equivalent to the particle-hole transformation allows
the partition function to be represented as a power
series in $e^{\mu\beta}$ or
$e^{\mu\beta}+e^{-\mu\beta}$ whose powers are ranging between
$[-n_xn_y,n_xn_y]$ or $[0,n_xn_y]$ respectively,
where the coefficients of the first expansion
are the canonical partition functions for a number of particles corresponding
to the index $n$, with $n=0$ being the half-filling partition function.
These coefficients are obtained by ensemble averaging
of the following
expansion of the statistical weight for each configuration
\begin{equation}\label{secondz}
\tilde z(\mu) =
e^{\mu n_xn_y\beta}
\sum_{n=-n_xn_y}^{n_xn_y}b_ne^{\mu\beta n}.
\end{equation}
or alternatively
\begin{equation}\label{firstz}
\tilde z(\mu)=
e^{\mu n_xn_y\beta}
\sum_{n=0}^{n_xn_y}a_n\left(e^{\mu\beta}+e^{-\mu\beta}\right)^n
\end{equation}
where the expansion coefficients in both these equations are obtained
from the eigenvalues of the matrix $P^-_{|\mu=0}$ [\ref{BarbourKlepfish}].
Alternatively, each of these expansions can be done around the updating
fugacity point, thus yielding for eqn.(\ref{firstz})
\begin{equation}\label{firstza}
\tilde z(\mu)=
e^{\mu n_xn_y\beta}
\sum_{n=0}^{n_xn_y}\tilde a_n\left(e^{\mu\beta}+e^{-\mu\beta}-e^{\mu_0\beta}
-e^{-\mu_0\beta}\right)^n
\end{equation}
with $\mu_0$ being the updating value of the chemical
potential. A similar expression can be obtained for eqn.(\ref{secondz}).
The coefficient of the zeroth power in (\ref{firstza}), normalised and
averaged over the ensemble of the HS field configurations, is
the average sign of the statistical weight calculated as
\begin{equation}\label{averagesign}
<{\rm sign}>={1\over {N_c}}\sum{{{\tilde z(\mu_0)}\over {|\tilde z(\mu_0)|}}}.
\end{equation}
where $N_c$ is the number of the HS field configurations accepted in the
Monte Carlo integration.
The statistical fluctuation in this coefficient reflects the
sign fluctuations.
The expansion for the partition function is then given by:
\begin{eqnarray}\label{pfexpansion}
Z(\mu) & = & e^{\mu n_xn_y\beta}
\sum_{n=0}^{n_xn_y} \left<{{\tilde a_n}\over {|\tilde z(\mu_0)|}}\right>
\nonumber\\
&\times & \left(e^{\mu\beta}+e^{-\mu\beta}-e^{\mu_0\beta}
-e^{-\mu_0\beta}\right)^n
\end{eqnarray}

When the average sign is near
unity, it is safe to assume that the lattice configurations
reflect accurately the quantum degrees of freedom. Following Blankenbecler
{\it et al.}
[\ref{Blankenbecler}] the diagonal matrix elements of the equal-time Green's
operator $G^\pm=(I+P^\pm)^{-1}$ accurately describe the fermion density
on a given configuration. In this regime
the adiabatic approximation, which is the
basis of the finite-temperature
algorithm, is valid. The situation differs strongly
when the average sign becomes small. We are in this case sampling
positive and negative $\tilde z(\mu_0)$ configurations with almost equal
probability since the acceptance criterion depends only on the
absolute value of $\tilde z(\mu_0)$.

In the simulations of the
HS fields the situation is different from the case
of fermions interacting with dynamical boson fields presented
in Ref.[\ref{Blankenbecler}]. The auxilary HS
fields do not have a kinetic
energy term in the bosonic action which would suppress their rapid
fluctuations and hence
recover the adiabaticity.
{}From the previous simulations on a $4^2$ lattice [\ref{BarbourKlepfish}]
we know that avoiding the sign problem, by updating at half-filling,
results in high uncontrolled fluctuations of the expansion
coefficients for the statistical weight, thus severely limiting
the range of validity of the expansion.
It is therefore important to obtain the partition
function for the widest range of $\mu_0$ and observe
the persistence of the hierarchy
of the expansion coefficients of $Z$. An error analysis
is required to establish the Gaussian distribution of the
simulated observables. We present in the following section
results of the bootstrap analysis [\ref{Hall}]
performed on our data for several
values of $\mu_0$.

\section{\bf Temperature and lattice-size dependence
of the Yang-Lee zeros}
The simulations were performed in the intermediate on-site repulsion
regime $U = 4t$ for $\beta=5,6,7.5$ on
lattices $4^2$, $6^2$, $8^2$ and for $\beta=5,6$ on a $10^2$ lattice.
The expansion coefficients given by eqn.(\ref{firstza})
are obtained with relatively
small errors and exhibit clear Gaussian distribution over the
ensemble. This behaviour was recorded for a wide range of
$\mu_0$ which makes
our simulations reliable in spite of the sign problem. In Fig.1 (a-c) we
present typical distributions of the first coefficients corresponding
to $n=1-7$ in eqn.(\ref{firstza}) (normalized with respect to the zeroth
power coefficient)
for $\beta= 5-7.5$ for different $\mu_0$.
The coefficients are
obtained using the bootstrap method on over 10000
configurations for $\beta=5$ increasing to over 30000
for $\beta=7.5$. In spite of
different values of the average sign in these simulations,
the coefficients
of the expansion (\ref{pfexpansion})  indicate good correspondence
between coefficients obtained with different values of
the update chemical potential $\mu_0$:
the normalized coefficients
taken from different $\mu_0$ values and equal power of the
expansion variable
correspond within the statistical error estimated using the bootstrap
analysis. (To compare these coefficients we had to shift
the expansion by $2\cosh {\mu_0\beta}$.)

We also performed a bootstrap analysis of the zeros
in the $\mu$ plane which shows clear Gaussian
distribution of their real and imaginary parts (see Fig\/.2).
In addition, we observe overlapping
results (i\/.e\/. same zeros) obtained with different values of $\mu_0$.
The distribution of Yang-Lee zeros in the complex
$\mu$-plane is presented in Fig\/.3(a-c) for the zeros nearest
to the real axis. We observe a gradual decrease of the
imaginary part as the lattice size increases.
The quantitative analysis of this behaviour is discussed
in the next section.

The critical domain can be identified by the behaviour of
the density of Yang-Lee zeros' in the positive half-plane of the
fugacity. We expect to find that this density is temperature and
volume dependent as the system approaches the phase transition.
If the temperature is much higher than the critical
temperature, the zeros stay far from the positive real axis
as it happens in the
high-temperature limit of the one-dimensional
Ising model ($T_c=0$) in which, for $\beta=0$, the points of singularity of
the free energy lie at fugacity value $-1$. As the temperature decreases
we expect the zeros to migrate to the positive half-plane with their
density, in this region, increasing with the system's volume.

Figures 4(a-c) show the number
$N(\theta)$ of zeros in the sector $(0,\theta)$ as a
function of the angle $\theta$. The zeros shown in these figures
are those presented in Fig\/.3(a-c) in the chemical potential
plane with other zeros lying further from the positive real half-axis
added in . We included only the zeros having absolute value
less than one which we are able to do because if $y_i$ is a zero
in the fugacity plane, so is
$1/{y_i}$. The errors are shown where they were estimated
using the bootstrap analysis (see Fig\/.2).

For $\beta=5$, even for the largest simulated lattice $10^2$, all the
zeros are in the negative half-plane. We notice a gradual movement
of the pattern of the zeros towards the smaller $\theta$ values
with an increasing density of the zeros near $\theta={\pi\over 2}$
which is due to the increase of the lattice size. However,
since no zeros are detected beyond this value of
$\theta$ we can assume that $\beta=5$ is too high a temperature
for a phase transition.

At $\beta=6$ we observe a qualitatively distinct behaviour
for the two biggest lattice sizes ($8^2$ and $10^2$). The low-lying
zeros are in the positive half-plane with
angular density estimated
as the slope of the linear interpolation between the points
on the graphs of $N(\theta)$ for a given lattice size. These
slopes increase with lattice size. Yet even the $10^2$
result extrapolated towards $\theta=0$ yields a vanishing density
of fugacity zeros on the positive real half-axis.

The $\beta=7.5$ results show the density of
zeros increasing with the simulation
volume
at small values of $\theta$. For the $8^2$ lattice the extrapolation
of $N(\theta)$ gives a non-vanishing slope at zero angle.
We interpret these results as signalling the existence of a criticality
for this temperature. To obtain the scaling of the
density of zeros near the positive half-axis we would have to
extend the simulation to larger lattices (work which is currently
in progress). On the basis of the existing results, we conclude
that the system described in our simulations approaches criticality
for the temperature corresponding to the lowest of our simulations,
namely $\beta=7.5$. We do not exclude the possibility that larger
lattice simulations may show a phase transition for an even slightly
higher temperature, as the $\beta=6$ results indicate.

\section{Finite size scaling and the singularity of the
density of states}

As a starting point for the
finite size analysis of the Yang-Lee singularities
we recall the scaling hypothesis for the partition function
singularities in the critical domain [\ref{ItzyksonPearsonZuber}].
Following this hypothesis, for a change of scale of the linear
dimension $L$
$$L\rightarrow {L\over \lambda}$$ we obtain a transformation
of the partition function
$$Z_L\rightarrow Z_{{L/\lambda}}$$
which, in the critical domain, satisfies
\begin{equation}\label{scalingone}
Z_L(\theta,{\tilde \mu})=
Z_{L/\lambda}(\theta_\lambda ,\tilde \mu_\lambda)
\end{equation}
where $\theta=({T\over {T_c}}-1)$, ${\tilde \mu}=(1-{\mu\over{\mu_c}})$
and $\theta_\lambda$ and  ${\tilde \mu}_\lambda$
are the values of $\theta$ and $\tilde \mu$ under the
rescaling transformation.
This equation is correct up to a multiplication
by an exponential stemming from the regular part of
the free energy.
In the scaling regime the following relationships hold
\begin{eqnarray}\label{scalingrelations}
\theta_\lambda&=&\lambda^{\alpha_\theta}\theta\nonumber\\
{\tilde \mu}_\lambda&=&\lambda^{\alpha_\mu}{\tilde \mu}\\
\lambda^dF_{\rm sing}
(\theta,{\tilde \mu})&=&F_{\rm sing}
(\theta_\lambda,{\tilde \mu}_\lambda)
\nonumber
\end{eqnarray}
where $\alpha_\theta$ and $\alpha_\mu$ are the scaling exponents
of the reduced temperature and reduced chemical potential respectively.
While the $\lambda$ dependence of $\theta_\lambda$ and $\tilde \mu_\lambda$
expressed in the first two equations of the
set (\ref{scalingrelations}) are correct for $\lambda\approx1$,
the third equation
(in which $F_{\rm sing}$ is the singular part of the free energy density),
follows in general from the conservation of the partition function
under the scaling transformation.
Here and in what follows we assume that this transformation does
not generate new couplings in the model.
Hence, the zeros of $Z_L$ will be the same as the zeros of
$Z_{{L/\lambda}}$.

Assume that
the simulation temperature is sufficiently low to
allow for a phase transition with
respect to the chemical potential.
In this case, the zeros of the grand canonical
partition function will correspond to the zeros of a
scaling function ${\cal Z}$
\begin{equation}\label{vanishing}
Z_{L/\lambda}(\theta \lambda^{\alpha_\theta},{\tilde \mu}\lambda^{\alpha_\mu})
={\cal Z}(\theta L^{\alpha_\theta},{\tilde \mu}L^{\alpha_\mu})=0.
\end{equation}
Solution of this equation determines the relationship between the
reduced chemical potential and the temperature through a function $f_i$
\begin{equation}\label{relationreduced}
{\tilde \mu_{i}}L^{\alpha_\mu}=f_i(\theta L^{\alpha_\theta})
\end{equation}
where $i$ is the index of the root of the partition function.
$\mu_c$ may vary with the temperature, and in
particular we can regard $\mu_c(T_{\rm simulation})$ as the critical
chemical potential at the temperature of simulation. Thus, for $\theta=0$
the previous equation becomes:
\begin{equation}\label{zerotheta}
{\tilde \mu}_iL^{\alpha_\mu}=f_i(0)
\end{equation}
which leads immediately to a determination of the exponent
$\alpha_\mu$ from the logarithmic scaling of the zeros
in the complex $\mu$ plane as
\begin{equation}\label{scalingmuspace}
\log{|\mu_L-\mu_c|_i}=-\alpha_\mu \log{L} + \log{f_i(0)}.
\end{equation}
The notation of the
last equation emphasizes the $L$-dependence of the chemical
potential $\mu$.

The above derivation of the finite size scaling
for the Yang-Lee zeros is a generalization of the
ideas developed for a two-dimensional Ising
model. In the present context, the essential difference
between the Hubbard and Ising models, is in the phase
boundary
$T_c(\mu)$. We do not expect this difference to significantly alter
the scaling relations with respect to the reduced chemical
potential. For the Ising
model the phase boundary in the $T-\mu$ plane is a straight
line $\mu=\mu_c=0$ (with the identification
of $\mu$ as the external magnetic field).
This line extends from $T=0$ to $T=T_c$, with the latter
being the critical temperature of a second-order phase transition.
Below $T_c$ the phase transition with respect to the
external field is first-order accompanied by
the characteristic two-phase coexistence.
We expect that for the Hubbard model the boundary will be given
by a line parametrised as $T_c(\mu)$. As this line is not
necessarily parallel to the $T$-axis, as one moves along either the
$T$ or $\mu$ directions, the phase boundary will be crossed.
Since the free energy is an analytic function on both sides
of the boundary, the limits of its derivatives are
independent of the path along which these limits are taken (as
long as the boundary is not crossed). Therefore, we would expect the
same order of singularity and phase transition
with respect to either temperature or chemical potential.

In what follows we assume a transition weaker than
a first-order one
(an assumption supported by simulation results of
Refs.[\ref{MoreoScalapinoDagotto}],
[\ref{Furukawa}]
for the 2D Hubbard
model) although even for a first-order phase transition
the finite-size scaling of the zeros is expected to be correct.

The critical exponent
$\alpha_\mu$ is related to the average particle number
via $\tilde \mu$ in the critical domain.
We have to determine the relationship between $\alpha_\mu$ and
$\delta$, where the latter is defined by
\begin{equation}\label{definedelta}
<n>\sim {\tilde \mu}^{1\over \delta}
\end{equation}
Following the real-space renormalization group treatment
of Ref.[\ref{ItzyksonPearsonZuber}] and assuming that the
change of scale $\lambda$ is a continuous parameter, the
exponent $\alpha_\theta$ is related to the critical
exponent $\nu$ of the
correlation length as $\alpha_\theta={1\over \nu}$.
This result follows from the scaling of the correlation
length $\xi$ as
\begin{equation}\label{xiscaling}
\xi(\theta_\lambda)={{\xi(\theta)}\over \lambda}.
\end{equation}
Thus, iterating the scaling of ${\theta}$ (Eq. \ref{scalingrelations})
and reaching the point
$\lambda\approx \theta^{-{1\over {\alpha_\theta}}}$
we obtain
\begin{equation}\label{xivsalphatheta}
\xi\sim |\theta|^{-{1\over {\alpha_\theta}}}=|\theta|^{-\nu}.
\end{equation}
(The absolute value in the last equation corresponds to the
points above and below $T_c$.)
Now recalling Eq.(\ref{scalingrelations})
and using the hyperscaling hypothesis ($F_{\rm sing}\sim \theta ^{\nu d}$)
we find:
\begin{equation}\label{hyperscaling}
F_{\rm sing}(\theta, \tilde \mu)=|\theta|^{\nu d}F_{\rm sing}
(\pm 1,{{\tilde \mu}\over
{|\theta|^{\nu\alpha_\mu}}})
\end{equation}
where $\theta_\lambda$ has been scaled to $\pm1$ and
${\tilde \mu}_\lambda$ expressed in terms of $\tilde \mu$ and $\theta$.
Differentiating this equation with respect to $\tilde \mu$
yields:
\begin{equation}\label{mtheta}
<n>_{\rm sing}=(-\theta)^{\nu(d-\alpha_\mu)}{{\partial F_{\rm sing}(X,Y)}
\over {\partial Y}}_{|X=\pm 1}
\end{equation}
which determines the critical exponent of the particle
density with respect to the reduced temperature as ${\nu(d-\alpha_\mu)}$.
Substituting $|\theta|\sim <n>_{\rm sing}^{1\over {\nu(d-\alpha_\mu)}}$
into
the argument $Y={{\tilde \mu}\over
{|\theta|^{\nu\alpha_\mu}}}$ and demanding the
hyperscaling $\theta$-dependence to be preserved in eqn.(\ref{hyperscaling})
we obtain:
\begin{equation}\label{mmu}
<n>\sim {\tilde \mu}^{{d-\alpha_\mu}\over {\alpha_\mu}}
\end{equation}
which defines the critical exponent
${1\over \delta}={{d-\alpha_\mu}\over {\alpha_\mu}}$
in terms of the scaling exponent $\alpha_\mu$ of the
Yang-Lee zeros.

Fig\/.5 presents the scaling of the imaginary part of the
$\mu$ zeros for different values of the temperature.
The linear regression slope of the
logarithm of the imaginary part
of the zeros plotted against the logarithm of
the inverse linear dimension of the
simulation volume, increases when the temperature
decreases from $\beta=5$ to $\beta=6$. The results of $\beta=7.5$
correspond to $\alpha_\mu=1.3$ within the errors of the zeros as
the simulation volume increases from $6^2$ to $8^2$.
As it is seen from Fig.\/3, we can trace zeros with similar
real part
($Re(\mu_1)\approx 0.7$ which is also consistent with the critical value
of the chemical potential given in Ref.[\ref{Furukawa}])
as the lattice size increases, which allows us
to examine only the scaling of the
imaginary part. Table 1 presents
the values of $\alpha_\mu$ and ${1\over \delta}$ obtained from this
figure.
\begin{center}
\begin{tabular}{|c|c|c|c|c|c|}
\hline
    \multicolumn{1}{|c|}{$\beta$} & \multicolumn{1}{c|}{$\alpha_\mu$}
&\multicolumn{1}{c|}{${1\over \delta}$}\\
\hline
 $5$ & $0.5\pm 0.05$
& $3.0\pm 0.4$ \\
\hline
 $6$ & $1.3\pm 0.2$
& $0.5\pm 0.2$\\
\hline
 $7.5$ & $1.3\pm 0.3$
& $0.5\pm 0.4$\\
\hline
\end{tabular}
\smallskip
\newline\noindent
Table 1 The finite size scaling exponents and the critical exponent
of particle-number singularity near criticality versus inverse temperature.
\end{center}
We also note that the location of the zeros corresponds
to the singularity of the electronic susceptibility $\chi$.
In Fig\/.6 we show $\chi$, given by
$\chi={{\partial <n>}\over {\partial \mu}}$,
as a function
of the chemical potential
on an $8^2$ lattice. The location of the peaks of the
susceptibility, rounded by the finite size effects, is
in good agreement with the distribution of the real part of
the Yang-Lee zeros in the complex $\mu$-plane (see Fig.3)
which is particularly evident in the $\beta=7.5$ simulations (Fig.4(c)). The
contribution of each zero to the susceptibility can be singled
out by expressing the free energy as:
\begin{equation}\label{singleout}
F=\sum_{i=1}^{2n_xn_y}(y-y_i)
\end{equation}
where $y$ is the fugacity variable and $y_i$ is the corresponding zero
of the partition function. The dotted lines on these plots correspond
to the contribution of the nearby zeros while the full polynomial
contribution is given by the solid lines. We see that the developing
singularities are indeed governed by the zeros closest to the real axis.
The sharpening of the singularity as the temperature decreases
is also in accordance with the dependence of the distribution of
the zeros on the temperature.

The singularities of the free energy and its derivative
with respect to the chemical potential, can be
related to the quasiparticle density of states.
To do this we assume that single particle excitations
accurately represent the spectrum of the system.
The relationship between the
average particle density and the density of states
$\rho(\omega)$ is given by
\begin{equation}\label{numberdensity}
<n>=\int^\infty_{0} d\omega
{{1\over {1+e^{\beta(\omega-\mu)}}}
\rho(\omega)}
\end{equation}
which in the low-temperature regime $\beta\rightarrow \infty $
becomes
\begin{equation}\label{lowtempapp}
<n>=<n>_{\rm reg}+<n>_{\rm sing}=\int_0^\mu d\omega {\rho(\omega)}
\end{equation}
Assuming $<n>_{\rm sing}$ dependence upon the chemical potential as
$<n>_{\rm sing}=(\mu-\mu_c)^{1/\delta}$
one obtains the zero-temperature
approximation for the density of states as the particle
susceptibility $\chi$, namely
\begin{equation}\label{densitysusc}
\chi_{\rm sing}={d<n>_{\rm sing}\over {d\mu}}=\rho_{\rm sing}(\mu)\propto
{1\over \delta}(\mu-\mu_c)^{{1\over \delta}-1}
\end{equation}
and hence the rate of divergence of the density of states.

As in the case of Lifshitz transitions the singularity of the
particle number is rounded at finite temperature. However, for
sufficiently low temperatures, the singularity of the density
of states remains manifest in the free energy, the average particle
density,
and particle susceptibility [\ref{Lifshitz}]. The regular
part of the density of states does not contribute to the criticality,
so we can concentrate on the singular part only. Consider
a behaviour of the type
$\rho_{\rm sing}(\omega)\propto (\omega-\mu_c)^{{1\over \delta}-1}$ for
$\omega > \mu_c$ and
$\rho_{\rm sing}(\omega)=0$ for $\omega < \mu_c$.
The integral (\ref{numberdensity}) will give in the region
$T<<(\mu-\mu_c)$ the
low-temperature
leading order behaviour for the particle
density:
\begin{equation}\label{finitetemperature}
<n>_{\rm sing}(\mu)\propto (\mu-\mu_c)^{1\over \delta}.
\end{equation}
with the value  $\delta$
for the particle number governed
by the divergence of the density of states (at
low temperatures)
in spite of the finite-temperature rounding of the singularity itself.
This rounding of the singularity is indeed
reflected in the difference between the values of $\alpha_\mu$ at
$\beta=5$ and $\beta=6$.

\section{Discussion and outlook}

We note that in our finite size scaling
analysis we do not include logarithmic corrections. In particular, these
corrections may
prove significant when taking into account the
fact that we are dealing with a two-dimensional system
in which the pattern of the phase transition is likely
to be of Kosterlitz-Thouless type [\ref{Kosterlitz}]. The
logarithmic corrections to the scaling laws have been proven
essential in a recent work of Kenna and Irving [\ref{kennairving}].
Inclusion of these corrections would allow us to obtain the
critical exponents with higher accuracy. However, such analysis
would require simulations on even larger lattices.

The linear fits for the logarithmic scaling and the critical
exponents obtained, are to be viewed as approximate values reflecting the
general behaviour of the Yang-Lee zeros as the temperature and lattice size are
varied. Although the bootstrap analysis provided us with
accurate estimates of the statistical error on the values of the
expansion coefficients and the Yang-Lee zeros, the small number of
zeros obtained with sufficient accuracy does not allow us to claim
higher precision for the critical exponents on the basis of more elaborate
fittings of the scaling behaviour.
The finite-size effects may still be significant, especially as the
simulation temperature decreases, thus affecting the scaling of the
Yang-Lee zeros with the system size. Larger lattice simulations
will therefore be required for an accurate evaluation of the critical
exponent for the particle density and the density of states.
Nevertheless, the onset of a singularity at finite temperature,
and its persistence as the lattice size increases, are evident.

The estimate of the critical exponent for the divergence rate
of the density of states of the quasiparticle excitation
spectrum
is particularly relevant to the high T$_c$ superconductivity
scenario
based on the van Hove singularities [\ref{Gofron}],[\ref{Newns}],
[\ref{Dagottobest}]. It is emphasized in Ref. [\ref{Gofron}]
that the logarithmic singularity of a two-dimensional electron gas
can, due to electronic correlations, turn into a power-law divergence
resulting in an extended saddle point at the lattice
momenta $(\pi,0)$ and $(0,\pi)$. In the case of the
density of states diverging as the $-{1\over 2}$ power, this singularity
leads to a weak-BCS mechanism for superconductivity with
$T_c$ set at $\approx 100$K [\ref{AbrikosovGofron}].
The extended saddle point behaviour has been confirmed
by experimental results and numerical calculations [\ref{Gofron}],
[\ref{Dessau}],[\ref{DagottoNazarenko}],[\ref{Bulut}].
In numerical simulations the extended saddle point can be traced
by a
reconstruction of the quasiparticle
spectral weight function from the simulation data on the Matsubara
Green's functions. This procedure requires
solving  an ill-posed inverse problem [\ref{WhiteVekic}],[\ref{Creffield}]
and is highly sensitive to the finite-doping simulation difficulties.
The results obtained so far, cannot provide a quantitative description
of the singularity.  The Yang-Lee analysis, however, provides a scheme
for a direct evaluation of the critical exponents.

We acknowledge stimulating
discussions with A. Bratkovsky, R. Kenna, P. Kornilovitch
and C. Creffield in the course of the work on this paper. This work
was supported by SERC grant GR/J 18675.

{\bf \noindent References}
\begin{enumerate}
\item\label{Blankenbecler}
{R. Blankenbecler, D.J. Scalapino and R.L. Sugar,
{\it Phys.Rev.} {\bf D40}, 2278 (1981).}
\item\label{Hirsch}
{J.E. Hirsch, {\it Phys.Rev.} {\bf B31}, 4403 (1985).}
\item\label{BarbourKlepfish}
{I.M. Barbour and E.G. Klepfish,
{\it Phys.Rev.} {\bf B46}, 469 (1992).}
\item\label{YangLee}
{C.N. Yang and T.D. Lee, {\it Phys. Rev.} {\bf 87}, 404 (1952);
\newline\noindent
T.D. Lee and C.N. Yang, {\it Phys. Rev.} {\bf 87}, 410 (1952).}
\item\label{Griffiths}{R.B. Griffiths, {\it Jour. Math. Phys.} {\bf 10}, 1559
(1969).}
\item\label{SuzukiFisher}{M. Suzuki and M.E. Fisher, {\it Jour. Math. Phys.}
{\bf 12}, 235 (1971).}
\item\label{DunlopNewman}{F. Dunlop and C.M. Newman, {\it Commun. Math Phys.}
{\bf 44}, 223 (1975).}
\item\label{SimonGriffiths}{B. Simon and R.B. Griffiths,
{\it Commun. Math Phys.} {\bf 33}, 145 (1973).}
\item\label{KCLee}{K-C. Lee, {\it Phys. Rev. Lett} {\bf 73}, 2801 (1994).}
\item\label{MerminWagner}
{N.D. Mermin and H. Wagner, {\it Phys. Rev. Lett} {\bf 17}, 1133
(1966).}
\item\label{ItzyksonPearsonZuber}{C. Itzykson, R.B. Pearson and
J.B. Zuber, {\it Nucl. Phys.}{\bf B220}, 415 (1983);\\
C. Itzykson and J-M. Drouffe, {\it Statistical Field Theory} {\bf v.1},
Cambridge University Press 1991.}
\item\label{Privman}
{M.L. Glasser, V. Privman and L.S. Schulman, {\it Phys. Rev.}{\bf B35}
(1987) 1841 .}
\item\label{Kenna}{R. Kenna and C.B. Lang, {\it Nucl. Phys.} {\bf B393},
461 (1993).}
\item\label{BarbourBellKlepfish}{I.M. Barbour, A.J. Bell and E.G. Klepfish,
{\it Nucl. Phys.} {\bf B389}, 285 (1993).}
\item\label{Lifshitz}{I.M. Lifshitz, {\it JETP} {\bf 38}, 1569 (1960).}
\item\label{Abrikosov}{A. A. Abrikosov, {\it Fundamentals of the
Theory of Metals} {\bf North-Holland} (1988).}
\item\label{Hall}{P. Hall, {\it The Bootstrap and Edgeworth expansion},
{\bf Springer} (1992).}
\item\label{White}
{S.R. White {\it et al.}, {\it Phys.Rev.} {\bf B40}, 506 (1989).}
\item\label{Hirschtr}
{J.E. Hirsch, {\it Phys. Rev.} {\bf B28}, 4059 (1983). }
\item\label{Suzuki}
{M. Suzuki, {\it Prog. Theor. Phys.} {\bf 56}, 1454 (1976).}
\item\label{MoreoScalapinoDagotto}{A. Moreo, D. Scalapino and E. Dagotto,
{\it Phys. Rev.}{\bf B43}, 11442 (1991).}
\item\label{Furukawa}{N. Furukawa and M. Imada, {\it J. Phys. Soc. Japan}
{\bf 61}, 3331 (1992).}
\item\label{Kosterlitz}{J. Kosterlitz and D. Thouless, {\it J. Phys.}
{\bf C6}, 1181 (1973);\\
J. Kosterlitz, {\it J. Phys.} {\bf C7}, 1046 (1974).}
\item\label{kennairving} {R. Kenna and A.C. Irving, {\it unpublished}.}
\item\label{Gofron}{K. Gofron {\it et al., Phys. Rev. Lett.} {\bf 73}, 3302
(1994).}
\item\label{Newns}{D.M. Newns, P.C. Pattnaik and C.C. Tsuei, {\it Phys.
Rev.} {\bf B43}, 3075 (1991);\\
D.M. Newns {\it et al.}, {\it Phys. Rev. Lett.} {\bf 24}, 1264 (1992);\\
D.M. Newns {\it et al.}, {\it Phys. Rev. Lett.} {\bf 73}, 1264 (1994).}
\item\label{Dagottobest}{E. Dagotto, A. Nazarenko and A. Moreo,
{\it Phys. Rev. Lett.} {\bf 74}, 310 (1995).}
\item\label{AbrikosovGofron}
{A.A. Abrikosov, J.C. Campuzano and K. Gofron, {\it Physica} (Amsterdam)
{\bf 214C}, 73 (1993).}
\item\label{Dessau}{D.S. Dessau {\it et al.},
{\it Phys. Rev. Lett.} {\bf 71}, 2781
(1993);\\
D.M. King {\it et al., Phys. Rev. Lett.} {\bf 73}, 3298 (1994);\\
P. Aebi {\it et al., Phys. Rev. Lett.} {\bf 72}, 2757
(1994).}
\item\label{DagottoNazarenko}{E. Dagotto, A. Nazarenko and M. Boninsegni,
{\it Phys. Rev. Lett.} {\bf 73}, 728 (1994).}
\item\label{Bulut}{N. Bulut,
D.J. Scalapino and S.R. White, {\it Phys. Rev. Lett.} {\bf 73},
748 (1994).}
\item\label{WhiteVekic}{S.R. White, {\it Phys. Rev.} {\bf B44}, 4670 (1991);\\
M. Veki\'c and S.R. White, {\it Phys. Rev.} {\bf B47}, 1160 (1993).}
\item\label{Creffield}{C.E. Creffield,
E.G. Klepfish, E.R. Pike and Sarben Sarkar, {\it unpublished}.}
\end{enumerate}
\bigskip
{\bf \noindent Figure Captions}\\
\\
{\bf Figure 1} \\
Bootstrap distribution of normalized
coefficients for expansion
(\ref{firstza}) at different update chemical potential $\mu_0$
for an $8^{2}$ lattice. The corresponding power of expansion
is indicated in the top figure. (a) $\beta = 5$,
(b) $\beta = 6$,
(c) $\beta = 7.5$. \\
\\
{\bf Figure 2} \\
Bootstrap distributions for the Yang-Lee zeros in the complex $\mu$
plane closest to
the real axis. (a) $10^{2}$ lattice
at $\beta = 5$,
(b) $10^{2}$ lattice
at $\beta = 6$,
(c) $8^{2}$ lattice
at $\beta = 7.5$. \\
\\
{\bf Figure 3} \\
Yang-Lee zeros in the complex $\mu$ plane closest to
the real axis. (a) $\beta = 5$,
(b) $\beta = 6$,
(c) $\beta = 7.5$. The corresponding lattice size is shown in
the top right-hand corner.\\
\\
{\bf Figure 4} \\
Angular distribution of the Yang-Lee
zeros in the complex fugacity plane
Error bars are drawn where estimated.
(a) $\beta = 5$,
(b) $\beta = 6$,
(c) $\beta = 7.5$.\\
\\
{\bf Figure 5} \\
Scaling of the imaginary part of $\mu_1$ ($Re(\mu_1)\approx
=0.7$) as a function of lattice size. $\alpha_mu$ indicates the
the fit of the logarithmic scaling.\\
\\
{\bf Figure 6} \\
Electronic
susceptibility as a function of chemical
potential for an $8^2$ lattice.
The solid line represents the contribution of all the $2n_xn_y$ zeros
and the dotted line the contribution of the six zeros nearest to
the real-$\mu$ axis.
(a) $\beta = 5$,
(b) $\beta = 6$,
(c) $\beta = 7.5$.

\end{document}